\documentclass[journal,onecolumn,12pt]{IEEEtran}

\usepackage[top=1in, bottom=1in, left=1in, right=1in]{geometry}
\usepackage{graphicx}

\usepackage{amsmath,amssymb}
\usepackage{dsfont}
\usepackage{cases}
\usepackage{bm}
\usepackage{amsfonts}
\usepackage[font=footnotesize]{caption}
\usepackage{indentfirst}
\usepackage{cite}
\usepackage{color}
\usepackage{caption}
\usepackage{algorithm}
\usepackage{algpseudocode}
\usepackage{booktabs}
\usepackage{subfigure}
\usepackage{multirow}
\usepackage{amsthm}
\usepackage{setspace}
  
\theoremstyle{remark}
\newtheorem{remark}{Remark}
\theoremstyle{}
\newtheorem{theorem}{Theorem}
\theoremstyle{}

\newtheorem{lemma}{Lemma}
\theoremstyle{}
\newtheorem{definition}{Definition}
\theoremstyle{remark}

\theoremstyle{definition}

\makeatletter
\newcommand{\tabcaption}{\def\@captype{table}\caption}
\makeatletter
 \allowdisplaybreaks[4]
\ifCLASSINFOpdf
\else
\fi

\definecolor{newcolor}{rgb}{0.5,0,1}
\definecolor{newcolor2}{rgb}{1,0,0.5}

\author{
Qifa Yan~\IEEEmembership{Member,~IEEE,} 
Xiaohu Tang~\IEEEmembership{Senior Member,~IEEE,}\\
Meixia Tao~\IEEEmembership{Fellow,~IEEE,} and
Qin Huang~\IEEEmembership{Senior Member,~IEEE}

\thanks{Q. Yan and X. Tang are with the Information Coding \& Transmission Key Lab of Sichuan Province, CSNMT Int. Coop. Res. Centre (MoST), Southwest Jiaotong University, Chengdu 611756, China(email: qifayan@swjtu.edu.cn, xhutang@swjtu.edu.cn). 

M. Tao is with the Department of Electronic Engineering, Shanghai Jiao Tong University, Shanghai 200240, China(email:mxtao@sjtu.edu.cn). 

Q. Huang is with the School of Electronic and Information Engineering, Beihang University, Beijing 100191, China (email:qinhuang@buaa.edu.cn).
}
}

\begin{document}

\title{A Fundamental   Tradeoff Among Storage, Computation, and Communication for Distributed Computing over Star Network}

\maketitle

\IEEEpeerreviewmaketitle

\begin{abstract}
Coded distributed computing can alleviate the communication load by leveraging the redundant storage and computation  resources with coding techniques in distributed computing. In this paper, we study a MapReduce-type distributed computing framework over star topological network, where all the workers exchange information through a common access point. The optimal tradeoff among the normalized number of the stored files (storage load), computed intermediate values (computation load) and transmitted bits in the uplink and downlink (communication loads)  are characterized.  A coded computing scheme is proposed to achieve the Pareto-optimal tradeoff surface, in which the access point only needs to perform simple chain coding between the signals it receives, and information-theretical bound matching the surface is also provided. 
\end{abstract}


\section{Introduction}
The rapid growth of computationally intensive applications on mobile devices has attracted much research interest in designing efficient distributed computing frameworks. One of the most important programing models for distributed computing is MapReduce \cite{MapReduce,Dryad}, which has been utilized to deal with computation tasks with data sizes as large as tens of terabytes.   

MapReduce framework allows to assign multiple computation tasks to distributed nodes, where each node only stores a subset of files. This is done by decomposing each function to be computed  into  a set of ``map" functions and  a ``reduce" function,  where  each map function can be computed from a batch of data, with the output called \emph{intermediate values} (IVs), while the computation of a ``reduce" function needs to collect  the IVs from all the data as inputs. The whole procedure  is composed of three phases, i.e., map, shuffle and reduce. In the map phase, each distributed node computes the map functions on its local file batch assigned by the server and generates   output IVs; in the shuffle  phase, the nodes exchange their computed IVs  to facitate each node to obtain the IVs needed by its assigned reduce functions;  in the reduce phase, each node computes its assigned reduce functions by decoding all the corresponding IVs.

Recently,  a \emph{coded distributed computing (CDC)} scheme was proposed by Li \emph{et al.} \cite{Li2018Tradeoff}, where the files are stored multiple times across the distributed nodes in the map phase. The IVs are also computed multiple times accordingly,  such that multicast opportunities are created for the shuffle phase. As a result, the communication load was reduced significantly compared to traditional uncoded scheme.   It was proved in \cite{Li2018Tradeoff} that the scheme achieves the optimal communication load for a given total storage requirements. Interestingly, the normalized number of files stored across the nodes was termed \emph{computation load} by Li \emph{et al}, because each node calculates all the IVs that can be obtained from the
data stored at that node in the model therein, no matter if these IVs are used or not in the subsequent phases. Subsequently, Ezzeldin \cite{Fragouli} and Yan \emph{et al} \cite{YanISWCS218, YanITW2018} found that some IVs are computed but not used in the model. For this reason, Yan \emph{et al} reformulated the problem as a tradeoff between storage, computation, and communication loads in \cite{Yan2022tradeoff}, which allows each node to choose any subset of IVs to compute from its stored files.

%

Some interesting works that extend CDC have been proposed, for example, the technique was combined with \emph{maximum distance separable} (MDS) code in matrix-vector multiplication tasks to resist stragglers in  \cite{Li2016unified}; stragglers with general functions are considered in \cite{YanPDA:straggler,YanStraggler:ISIT};  the optimal resource allocations are considered in  \cite{Qian2017How}; \cite{Attia2016iterative1,Attia2016iterative2,Adel2018} investigated the  iterative procedures of data computing and shuffling; \cite{Song1} studied the case when each node has been randomly allocated files; \cite{Song2} investigated the case with random connectivity between nodes. 

 The coded distributed computing technique is extended to wireless distributed computing \cite{Li2018Wireless,Lampiris2018ISIT}, where the computation is typically carried out by the wireless devices.  Due to the decentralized natural of the wireless networks, the nodes in wireless networks  normally need a central \emph{Access Point}  (AP) to exchange data, which leads to  uplink and downlink communications. For example, smart-phone end users typically communicate with each other through a base station in cellular
networks, which operates in a star network.
 In \cite{Li2017Egdge} and \cite{Li2017Framework}, Li \emph{et al.} investigated distributed computing in a wireless network where the nodes performs data shuffling through an AP.  The optimal storage-communication tradeoff was characterized for both uplink and downlink transmissions.  

In this paper, following the conventions of Ezzeldin \cite{Fragouli} and Yan \emph{et al} \cite{Yan2022tradeoff}, we investigate a distributed computing system with star network, where all nodes exchange IVs through an AP, but each node is allowed to choose any arbitrary subset of IVs to compute from its stored files.  In particular, in addition to the storage and computation loads as considered in  \cite{Yan2022tradeoff}, the communication load includes both upload and download. \emph{The main contribution of this paper is the characterization of the Pareto-optimal surface in the storage-computation-upload-download space for distributed computing over star network.} The idea is to form the same multicast signals as in CDC scheme but compute less IVs by ignoring  the un-used IVs in the map phase in the uplink, and combine them through a simple chain coding to form the downlink signals at the  AP. It turns out that, for any given storage-computation pair, both the optimal upload and download communication costs can be simultaneously achieved by a coded computing scheme that oriented from CDC. The information-theoretical bound matching the Pareto-optimal surface is also presented.

\emph{Paper Organization:}
 Section~\ref{sec:model} presents the system model.
 Section~\ref{sec:main:result}  summarizes  the main results.  Section \ref{sec:achieve} presents the coded computing scheme that achieves the optimal surface, and Section~\ref{sec:converse} provides information-theoretical bound. Finally, Section~\ref{sec:conclusion} concludes the paper.

\textit{Notations:} Let $\mathbb{N}^+$ be the set of positive integers, and $\mathbb{F}_2$ be the binary field. For $m,n\in\mathbb{N}^+$, denote the $n$-dimensional vector space over   $\mathbb{F}_2$ by $\mathbb{F}_{2}^n$, and the integer set $\{1,\ldots,n\}$ by $[n]$. 
If $m<n$, we use $[m:n]$ to denote the set $\{m,m+1,\ldots,n\}$.   We also use   interval notations, e.g., $[a,b]\triangleq\{x:a\leq x\leq b\}$ and $[a,b)\triangleq\{x:a\leq x<b\}$ for real numbers $a,b$ such that $a<b$. The bitwise exclusive OR (XOR) operation is denoted by $\oplus$.
For sets we use upper case calligraphic font, e.g., $\mathcal{A}$, and for collections (sets of sets) we use upper case  Greek letters with bold font, e.g., $\boldsymbol{\Omega}$.  We denote a point in two or three dimensional Euclidean space by an upper case letter.
%
 A  line segment with end points $A_1,A_2$ or a line through  the points $A_1,A_2$ is denoted by $A_1A_2$.
  A triangle with vertices $A_1,A_2, A_3$ is denoted by  $\triangle   A_1A_2A_3$. A trapezoid with the four edges $A_1A_2$,
  $A_2A_3$, $A_3A_4$, and $A_4A_1$,
   where $A_1A_2$ is parallel to $A_3A_4$, is denoted  by $\boxminus A_1A_2A_3A_4$.  Let $\mathcal{F}$ be a set of facets, if the facets in $\mathcal{F}$ form  a continuous surface, then we refer to this surface simply as $\mathcal{F}$.

\section{System Model}\label{sec:model}
\begin{figure}[t]
                \centering
               \includegraphics[scale=0.3]{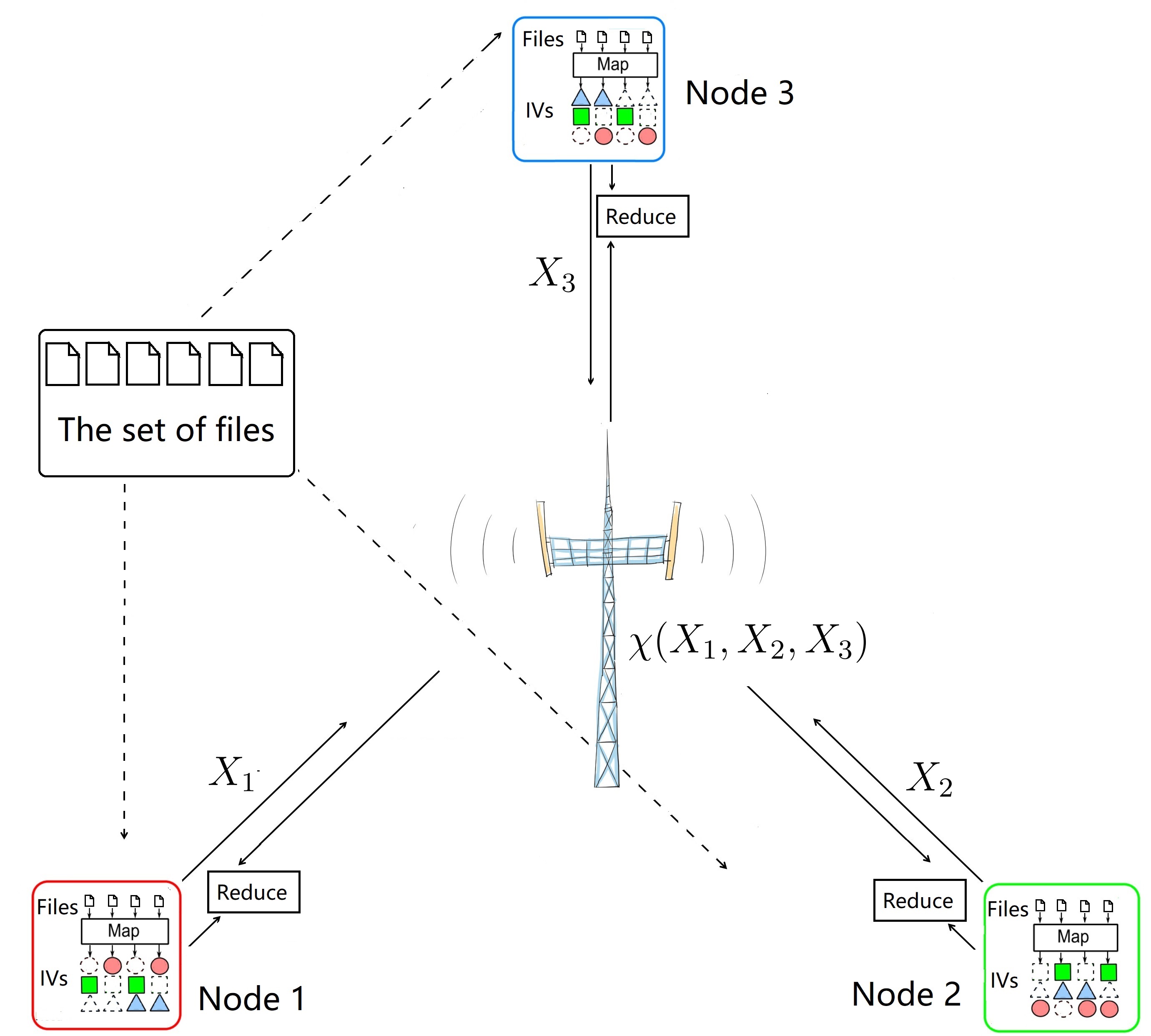}
                \caption{A Distributed Computing System with Star Network}\label{fig:system:model}
\end{figure}

Let $K,N,W,U,V$ be given positive integers. Consider a star network consisting of $K$ distributed computing nodes $\{1,\ldots,K\}$ that can communicate with each other through  a common AP, as illustrated in Fig. \ref{fig:system:model}.  Each of the $K$ nodes can transmit signals to the AP through an uplink channel, while the AP can broadcast signals to all the $K$ nodes via a downlink channel.

Each of the $K$ nodes aims to compute an individual function from a set of $N$
 files,
\begin{IEEEeqnarray}{c}
\mathcal{W}=\{w_1,\ldots,w_N\},\quad w_n\in \mathbb{F}_2^W,\forall~n\in[N],\notag
 \end{IEEEeqnarray}
 each of size $W$ bits.
Node $k$ aims to compute an output function
 \begin{IEEEeqnarray}{c}
 \phi_k\colon \mathbb{F}_2^{NW}\rightarrow \mathbb{F}_2^U,\notag
 \end{IEEEeqnarray}
 which maps all the files to a bit stream
 \begin{IEEEeqnarray}{c}
 u_k=\phi_k(w_1,\ldots,w_N)\in\mathbb{F}_2^U\notag
 \end{IEEEeqnarray}
  of length $U$.
 Assume that each output function $\phi_k$ decomposes as:
\begin{IEEEeqnarray}{c}
\phi_k(w_1,\ldots,w_N)=h_k(f_{k,1}(w_1),\ldots,f_{k,N}(w_N)),\label{eq:output}
\end{IEEEeqnarray}
where
\begin{itemize}
\item Each \emph{``map" function} $f_{k,n}$ is of the form
\begin{IEEEeqnarray}{c}
f_{k,n}: \mathbb{F}_2^W\rightarrow \mathbb{F}_2^V,\notag
\end{IEEEeqnarray}
and maps the file $w_n$ into the IV
 \begin{IEEEeqnarray}{c}
 v_{k,n}\triangleq f_{k,n}(w_n)\in \mathbb{F}_2^V.\notag
 \end{IEEEeqnarray}
  \item The \emph{``reduce" function} $h_k$ is of the form
\begin{IEEEeqnarray}{c}
  h_k: \mathbb{F}_2^{NV}\rightarrow \mathbb{F}_2^U, \notag
\end{IEEEeqnarray}
 and  maps the IVs
  \begin{IEEEeqnarray}{c}
  \mathcal{V}_k\triangleq\{v_{k,n}:n\in [N]\}\notag
  \end{IEEEeqnarray}
   into the output stream
   \begin{IEEEeqnarray}{c}
   u_k=h_k(v_{k,1},\ldots,v_{k,N}).\notag
   \end{IEEEeqnarray}
   \end{itemize}

Notice that  one trivial decompositon is that,  the map functions  are identity functions and the reduce functions are the output functions, i.e., $g_{k,n}(w_n)=w_n,$ and $h_k=\phi_k,~\forall~n\in [N],~k\in[K]$. But in practice,    many output functions can be decomposed such that the main computation load is dominated  by the map functions.   For example, in federated learning, it typically needs to  collect the sum of the gradients over all data blocks, where the map functions are used  to compute the gradients of the loss functions over  a data block, while the  reduce function is the sum operation.

The described structure  of the output functions $\phi_1,\ldots, \phi_K$, allows the nodes to perform their computation in the following three-phase procedure.

1) \textbf{Map Phase:}  Each node $k\in[K]$ chooses to  store a subset  of files $\mathcal{M}_k\subseteq \mathcal{W}$.
For each file $w_n\in\mathcal{M}_k$,  node $k$ computes a subset of IVs
\begin{IEEEeqnarray}{c}
\mathcal{C}_{k,n}=\{v_{q,n}:q\in \mathcal{Z}_{k,n}\},\notag
\end{IEEEeqnarray}
where $\mathcal{Z}_{k,n}\subseteq[K]$. Denote the set of IVs computed at node $k$ by $\mathcal{C}_k$, i.e.,
\begin{IEEEeqnarray}{c}
    \mathcal{C}_k\triangleq \bigcup_{n:w_n\in\mathcal{M}_k}\mathcal{C}_{k,n}.\label{defeqn:Ck}
\end{IEEEeqnarray}

 2) \textbf{Shuffle Phase:}
The  $K$ nodes exchange some of their computed IVs through the AP  via \emph{upload} and \emph{download} sub-phases:

In the \emph{upload sub-phase}, each  node $k$ generates a coded signal
\begin{IEEEeqnarray}{c}
       X_k=\varphi_k\left(\mathcal{C}_k\right)\notag
\end{IEEEeqnarray}
of some length  $l_k\in\mathbb{N}$ and sends it to the AP, using a function
\begin{IEEEeqnarray}{c}
       \varphi_k: \mathbb{F}_2^{|\mathcal{C}_k|V}\rightarrow \mathbb{F}_2^{l_k}.\notag
\end{IEEEeqnarray}

In the \emph{download sub-phase}, receiving all the signals $\{X_1,\ldots,X_K\}$,  the AP generates a signal 
\begin{IEEEeqnarray}{c}
X=\chi(X_1,X_2,\ldots,X_K)\label{eqn:X}
\end{IEEEeqnarray}
 of length $l\in\mathbb{N}$, and broadcasts it to all nodes, where the encoding function is
\begin{IEEEeqnarray}{c}
\chi: \mathbb{F}_2^{l_1+l_2+\ldots+l_K}\rightarrow \mathbb{F}_2^l.\notag
\end{IEEEeqnarray}

3) \textbf{Reduce Phase:} Using the received signal $X$ broadcast from the AP in the shuffle phase and its own IVs $\mathcal{C}_k$ computed locally in the map phase,  each node $k$ now computes the IVs
 \begin{IEEEeqnarray}{c}
       (v_{k,1},\ldots,v_{k,N})=\psi_k\left(X,\mathcal{C}_k\right),\label{eqn:adjust1b}
\end{IEEEeqnarray}
for some function
  \begin{IEEEeqnarray}{c}
       \psi_k: \mathbb{F}_2^{l}\times \mathbb{F}_2^{|\mathcal{C}_k|V}\rightarrow \mathbb{F}_2^{NV}.\notag
\end{IEEEeqnarray}
    Finally, it   computes
 \begin{IEEEeqnarray}{c}
       u_k=h_k(v_{k,1},\ldots,v_{k,N}).\label{eqn:reducefunction}
\end{IEEEeqnarray}

To measure the storage, computation, and communication costs of the described procedure, following the convention in \cite{Yan2022tradeoff}, we introduce the following definitions.
\begin{definition}[Storage load]
\emph{Storage load} $r$ is defined as the total number of files stored across the $K$ nodes normalized by the total number of files $N$:
\begin{IEEEeqnarray}{c}
r\triangleq\frac{\sum_{k=1}^K|\mathcal{M}_k|}{N}.\label{defeqn:r}
\end{IEEEeqnarray}
\end{definition}

\begin{definition}[Computation load] \emph{Computation load} $c$ is defined as the total number of map functions computed across the  $K$ nodes,  normalized by the total number of map functions $NK$:
  \begin{IEEEeqnarray}{c}
    c \triangleq\frac{\sum_{k=1}^K|\mathcal{C}_k|}{NK}.\label{defeqn:c}
  \end{IEEEeqnarray}
\end{definition}

\begin{definition}[Communication Load]\label{def:comload}  The communication load is characterized by the tuple $(L,D)$, where  $L$ (resp. $D$) is the upload (resp. download)  defined as the total number of the bits sent by the $K$ nodes (resp. AP)  during the upload (resp. download) sub-phase, normalized by the total length of all intermediate values $NKV$:
  \begin{IEEEeqnarray}{c}
    L\triangleq \frac{\sum_{k=1}^K l_k}{NKV},\quad D\triangleq\frac{l}{NKV}.\notag
  \end{IEEEeqnarray}
\end{definition}


\begin{remark}[Nontrivial Regime]\label{remark:nontrivial}
 In general, the non-trivial regime in our setup is
\begin{subequations} \label{eq:interesting}
\begin{IEEEeqnarray}{c}
1\leq c\leq r\leq K,\label{eq:interesting_a}\\
  0 \leq D\leq L \leq 1- \frac{r}{K}.\label{eq:interesting_b}
\end{IEEEeqnarray}
\end{subequations}
For completeness, we justify them by the following observations. 
\begin{itemize}
\item \emph{Justification of \eqref{eq:interesting_a}:} Since each IV needs to be computed
at least once somewhere, we have $c\geq 1$. Moreover, the definition of
$\mathcal{C}_k$ in  \eqref{defeqn:Ck} implies that
$|\mathcal{C}_k|\leq|\mathcal{M}_k|K$, and thus by \eqref{defeqn:r} and
\eqref{defeqn:c}, $c\leq r$. Finally, the regime $r> K$ is not interesting, because in this case each node stores all the files, and can thus locally compute all the IVs required to compute its output function. In this case, $c\geq 1, D\geq 0$ and $L\geq 0,$ can be arbitrary.
\item  \emph{Justification of \eqref{eq:interesting_b}:} $D\geq 0$ is trivial. By \eqref{eqn:X}, as the down-link signal $X$ is created from the upload signals $X_1,\ldots,X_K$,  $D=L$ is sufficient to communicate all the received information. Finally, each node $k$ can trivially compute
$|\mathcal{M}_k|$ of its desired IVs  locally and thus only needs to
receive $N-|\mathcal{M}_k|$ IVs from other nodes. Thus, such an uncoded manner requires an upload of $L=\frac{\sum_{k=1}^K(N-|\mathcal{M}_k|)V}{NKV}=1-\frac{r}{K}$.
\end{itemize}
In the trivial case that the AP simply  forwards all the receiving signals, i.e., $X=(X_1,\ldots,X_K)$, then $D=L$, and the model degrades to the distributed model without the AP as in \cite{Yan2022tradeoff}, where the non-trivial region on the triple $(r,c,L)$ was $1\leq c\leq r\leq K, 0\leq L\leq 1-\frac{r}{K}$. 
\end{remark}
  
\begin{definition}[Fundamental SCC Region] A Storage-Computation-Communication\footnote{The communication load includes both upload and download.} (SCC) quadruple $(r,c,L,D)$ satisfying \eqref{eq:interesting} is achievable if for any
$\epsilon>0$ and sufficiently large $N,W,V$, there exist   map, shuffle, and
reduce procedures with  storage load, computation load, upload and download less than $r+\epsilon$, $c+\epsilon$, $L+\epsilon$ and $D+\epsilon$, respectively. The fundamental SCC region is defined as the set of all feasible SCC quadruple:
\begin{IEEEeqnarray}{c}
\mathcal{R}=\{(r,c,L,D): (r,c,L,D)~\mbox{is feasible}\}.\notag
\end{IEEEeqnarray}
\end{definition}

\begin{definition}[Optimal Tradeoff Surface]
An SCC quadruple $(r,c,L,D)$ is called \emph{Pareto-optimal} if it is feasible and if no feasible  SCC quadruple $(r',c',L',D')$ exists so that  $r' \leq r,c'\leq  c, L'\leq  L$ and $D\leq D'$ with one or more of the inequalities being strict. The set of all Pareto-optimal SCC quadruples is defined as the optimal tradeoff surface:
\begin{IEEEeqnarray}{c}
\mathcal{O}\triangleq\{(r,c,L,D):(r,c,L,D)~\mbox{is Pareto-optimal}\}.\notag
\end{IEEEeqnarray}
\end{definition}
The goal of this paper is to characterize the fundamental SCC region $\mathcal{R}$ and the optimal tradeoff surface $\mathcal{O}$ in our setup. 

\section{Main Results}\label{sec:main:result}

Before we present the main theorem, let us provide a toy example to illustrate the key idea of the proposed achievable scheme.

\subsection{An Toy Example for Achievable Scheme }
\begin{figure}[t] \centering
\subfigure[] { \label{fig:a}
\includegraphics[scale=0.25]{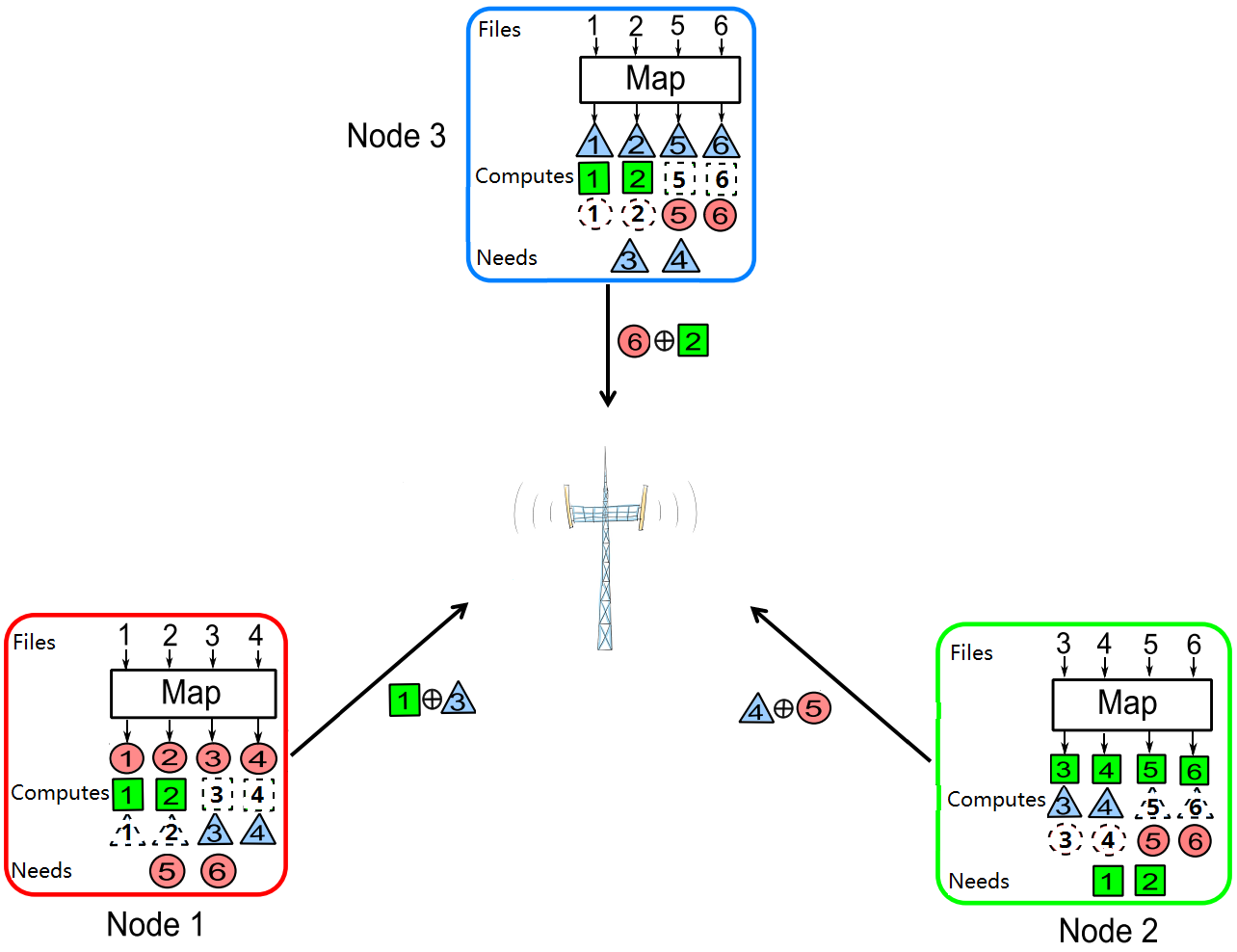}
}
\subfigure[] { \label{fig:b}
\includegraphics[scale=0.25]{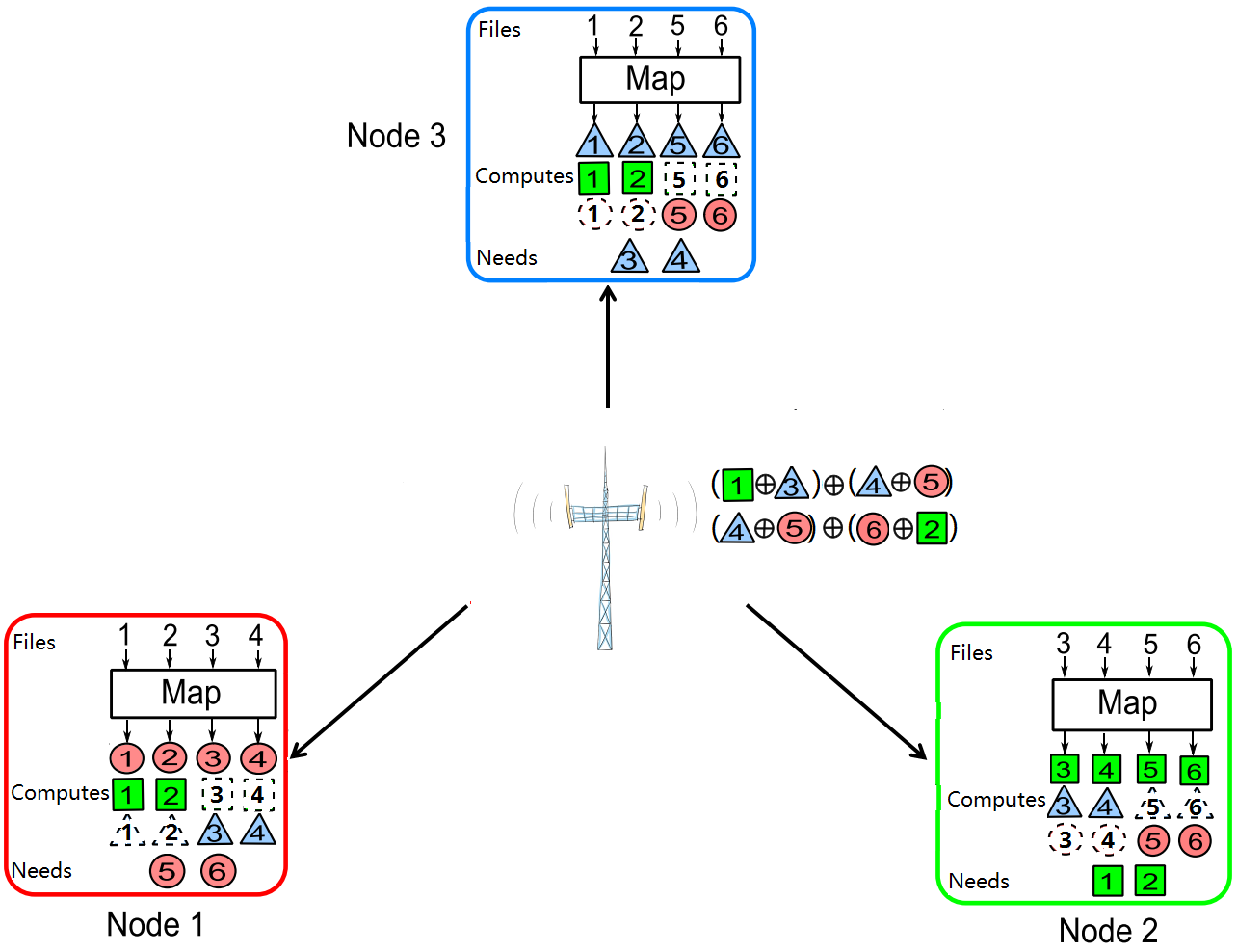}
}
\caption{Illustration of the CDC for star network: (a) Uplink (b) Downlink }
\label{fig:example}
\end{figure}

 Consider the case, where there aorre $K=3$ nodes and $N=6$ files.  Each node wants to compute an individual function from the $N=6$ files  as in \eqref{eq:output}. Fig. \ref{fig:example} illustrates the strategy achieving the Pareto-optimal point $(r,c,L,D)=(2,\frac{4}{3},\frac{1}{6},\frac{1}{9})$, where the uplink and downlink transmissions  are illustrated in Fig.\ref{fig:a} and \ref{fig:b}, respectively.

In Fig. \ref{fig:example}, the three nodes are denoted by three boxes with red, green and blue edges respectively. The top-most lines in each of the three boxes indicate the files stored at the node. The rectangle below this line indicates the map functions at the node.  
 The computed IVs are depicted below the rectangle, where red
circles, green squares, and blue triangles indicate IVs $\{v_{1,1},\cdots,v_{1,6}\}$, 
$\{v_{2,1}, \cdots , v_{2,6}\}$, and  $\{v_{3,1}, \cdots , v_{3,6}\}$,  respectively. The
dashed circles/squares/triangles stand for the IVs that are not
computed from the stored files. The last line of each box
indicates the IVs that the node needs to learn during the
shuffle phase.  

The $N=6$ files 
\begin{IEEEeqnarray}{c}
\mathcal{W}=\{w_1,w_2,w_3,w_4,w_5,w_6\}\notag
\end{IEEEeqnarray}
are partitioned into ${K\choose r}=3$ batches, i.e., 
$
\{w_1,w_2\},\{w_3,w_4\},
\{w_5,w_6\}$. 
In the map phase, the files $\{w_1,w_2\}$ are simultaneously stored at nodes $1$ and $3$; the files $\{w_3,w_4\}$  at  nodes $1$ and $2$; and the files $\{w_5,w_6\}$ at nodes $2$ and $3$. For each node, the computed IVs can be classified into two types:  the IVs that will be used by its own reduce function (the first line below the ``map" rectangle) and the IVs that will be used for transmission or decoding (the second and third lines below the ``map" rectangle).

In the shuffle phase, during the upload sub-phase, each node creates a coded signal by XORing two IVs  and sends it to the AP as illustrated in Fig. \ref{fig:a}, i.e., Nodes $1$, $2$ and $3$ sends coded IVs $v_{1,1}\oplus v_{3,3}, v_{3,4}\oplus v_{2,5}$ and $v_{6,2}\oplus v_{2,1}$, respectively; during the download sub-phase, the AP combines the three received signals by a simple chain coding, i.e., the two downlink signals are formed by XORing the signals from nodes $1$ and $2$, and the signals from $2$ and $3$, respectively. The combined signals are sent to all the three nodes. 

In the reduce phase, for each node, since the two chain coded signals involve a coded signal transmitted by itself, the node can decode the two coded signals from the other two nodes. Moreover, from each of the coded singals, the node can further decode an IVs it needs, by XORing the coding signal with one of its computed IV. For example, Node $1$ first decodes the two signals $v_{3,4}\oplus v_{2,5}$ and $v_{2,6}\oplus v_{1,2}$, then it can further decodes the IVs $v_{2,5}$ and $v_{2,6}$, since the IVs $v_{3,4}$ and $v_{1,2}$ have been computed locally. Finally, each node collects all IVs for its assigned reduce function, and computes the final output.

\subsection{Fundamental SCC Region and Optimal Tradeoff Surface}

For each $i\in[K]$, define two SCC quadrules
\begin{IEEEeqnarray}{l}
P_i\triangleq\left(i,i\left(1-\frac{i-1}{K}\right),\frac{1}{i}\left(1-\frac{i}{K}\right),\frac{1}{i+1}\left(1-\frac{i}{K}\right)\right),\notag\\
Q_i\triangleq\left(i,i,\frac{1}{i}\left(1-\frac{i}{K}\right),\frac{1}{i+1}\left(1-\frac{i}{K}\right)\right).\notag
\end{IEEEeqnarray}
In the following,  we will use $P_i^{\rm u},Q_i^{\rm u}, P_i^{\rm d}, Q_i^{\rm d}$ to denote the projections of $P_i,Q_i$ into the uplink and downlink SCC subspaces\footnote{In this paper, we will refer $r$-$c$-$L$ subspace as the uplink SCC subspace, and the $r$-$c$-$D$ subspace the downlink SCC subspace. The superscripts ``$\rm u$" and ``$\rm d$" indicate ``uplink" and ``downlink", respectively.}, i.e., 
\begin{IEEEeqnarray}{rCl}
P_i^{\rm u}&\triangleq&\left(i,i\left(1-\frac{i-1}{K}\right),\frac{1}{i}\left(1-\frac{i}{K}\right)\right),\notag\\
 Q_i^{\rm u}&\triangleq&\left(i,i,\frac{1}{i}\left(1-\frac{i}{K}\right)\right),\notag\\
P_i^{\rm d}&\triangleq&\left(i,i\left(1-\frac{i-1}{K}\right),\frac{1}{i+1}\left(1-\frac{i}{K}\right)\right),\IEEEeqnarraynumspace\label{def:Pd}\\
 Q_i^{\rm d}&\triangleq&\left(i,i,\frac{1}{i+1}\left(1-\frac{i}{K}\right)\right).\notag
\end{IEEEeqnarray}
%
The main result of this paper is summarized in the following theorem, where the proofs are provided in the following sections.
\begin{theorem}
The fundamental SCC region $\mathcal{R}$ is given by 
\begin{IEEEeqnarray}{rCl}
\mathcal{R}&=&\Big\{(r,c,L,D): 1\leq c\leq r\leq K,  \,L^*(r,c)\leq L\leq1-\frac{r}{K}, \, D^*(r,c)\leq D\leq L\Big\},\notag
\end{IEEEeqnarray}
where $L^*(r,c)$ is a function such that  $\{(r,c,L^*(r,c)):1\leq c\leq r\leq K\}$  forms the surface 
\begin{IEEEeqnarray}{rCl}
\mathcal{F}^{\rm u}&\triangleq&\triangle P_1^{\rm u}P_2^{\rm u}Q_2^{\rm u}\cup \mathop\cup_{i=2}^{K-1} \triangle P_{i-1}^{\rm u}P_i^{\rm u}P_K^{\rm u}\cup\mathop\cup_{i=2}^{K-1}\boxminus P_i^{\rm u}Q_i^{\rm u}Q_{i+1}^{\rm u}P_{i+1}^{\rm u} \notag
\end{IEEEeqnarray}
in the uplink SCC subspace, 
and $D^*(r,c)$ is a function such that  $\{(r,c,D^*(r,c)):1\leq c\leq r\leq K\}$  forms the surface
\begin{IEEEeqnarray}{rCl}
\mathcal{F}^{\rm d}&\triangleq&\triangle P_1^{\rm d}P_2^{\rm d}Q_2^{\rm d}\cup \mathop\cup_{i=2}^{K-1} \triangle P_{i-1}^{\rm d}P_i^{\rm d}P_K^{\rm d}\cup\mathop\cup_{i=2}^{K-1}\boxminus P_i^{\rm d}Q_i^{\rm d}Q_{i+1}^{\rm d}P_{i+1}^{\rm d} \notag
\end{IEEEeqnarray}
in the downlink SCC subspace. 
The optimal tradeoff surface is given by
\begin{IEEEeqnarray}{rCl}
\mathcal{O}&=&\mathop\cup_{i=2}^{K-1}\{\theta_1 P_{i-1}+\theta_2 P_{i}+\theta_3 P_K: \theta_1,\theta_2,\theta_3\in[0,1], \theta_1+\theta_2+\theta_3=1\}. \label{surfaceO}
\end{IEEEeqnarray}
\end{theorem}

\begin{figure*}[t]
 	\centering
 	\includegraphics[width=1\textwidth]{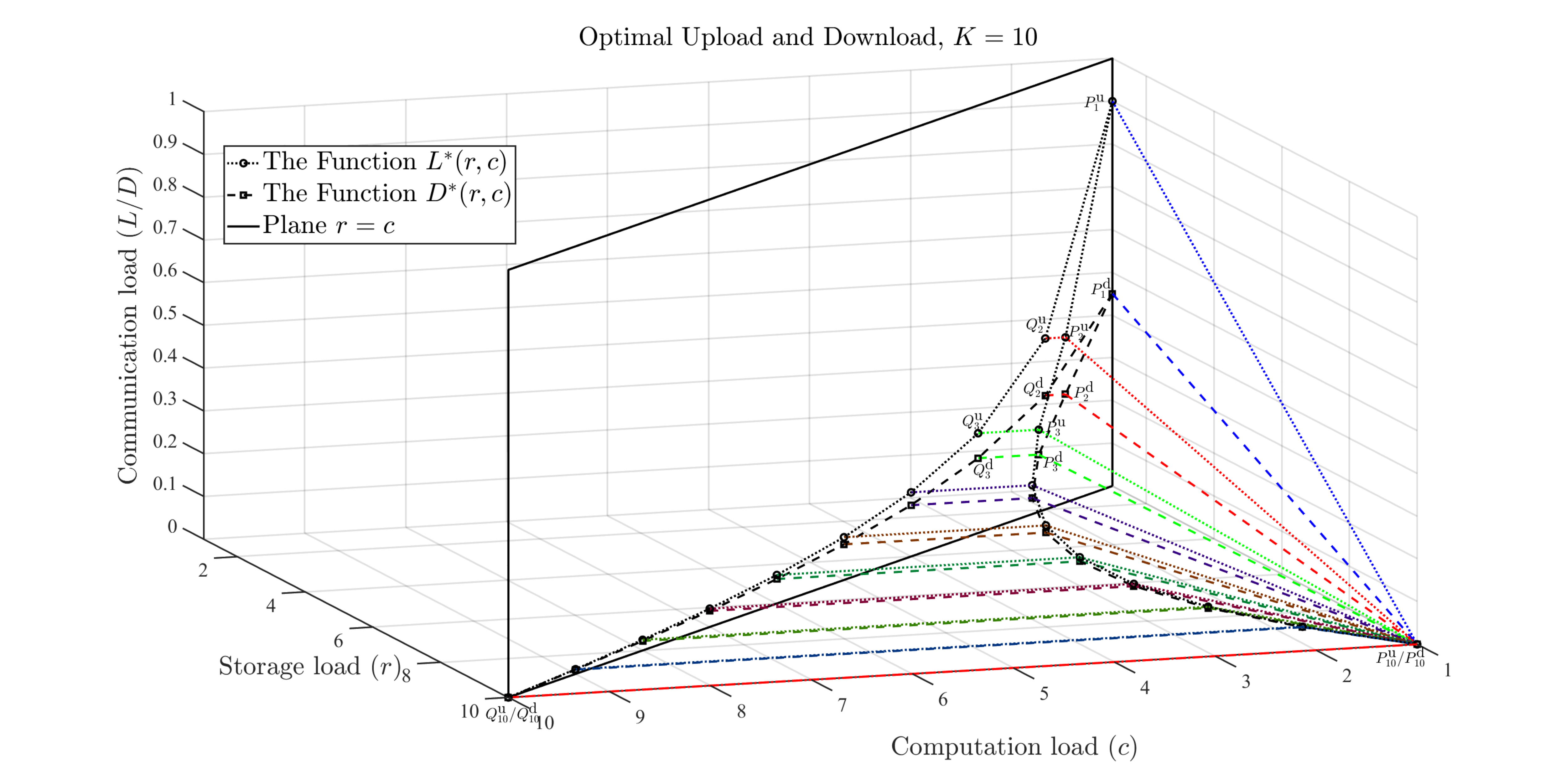}
 	\caption{The functions $L^*(r,c)$ and $D^*(r,c)$.
 		 }\label{fig:UpDown}
 \end{figure*}

In Fig. \ref{fig:UpDown}, the functions $L^*(r,c)$ and $D^*(r,c)$ are ploted for $K=10$ nodes.   Notice that, by setting $r=c$, we recover the optimal upload and download as investigated in  \cite{Li2017Framework}\footnote{The measurement of  communication load is up to a scalar ``$K$" in  \cite{Li2017Framework} compared Definition \ref{def:comload}, and a slightly difference in assumption in \cite{Li2017Framework} is that each node has a fixed storage load.}, i.e., 
\begin{enumerate}
\item the optimal upload  for given storage is given by
\begin{IEEEeqnarray}{c}
L^*(r)\triangleq {\rm{Conv}}\Big\{\frac{1}{r}\Big(1-\frac{r}{K}\Big)\Big\},\notag
\end{IEEEeqnarray}
which corresponds the curve formed by the line segments $Q_1^{\rm{u}}Q_2^{\rm{u}}, Q_2^{\rm{u}}Q_3^{\rm{u}},\ldots,Q_{K-1}^{\rm{u}}Q_K^{\rm{u}}$.
\item the optimal download for given storage is given by
\begin{IEEEeqnarray}{c}
D^*(r)\triangleq {\rm{Conv}}\Big\{\frac{1}{r+1}\Big(1-\frac{r}{K}\Big)\Big\},\notag
\end{IEEEeqnarray}
which corresponds to the curve formed by the line segments  $Q_1^{\rm d}Q_2^{\rm d}, Q_2^{\rm d}Q_3^{\rm d},\ldots,Q_{K-1}^{\rm d}Q_K^{\rm d}$.
\end{enumerate}

Observe that, the line segments $Q_i^{\rm u}P_i^{\rm u}$ in the uplink SCC space and $Q_i^{\rm d}P_i^{\rm d}$ in the downlink space ($i=2,3,\ldots,K$) are  parellel to the $c$-axis, which indicate  that the computation load can be saved to achieve $L^*(r)$.  The length of the line segments indicates the  amount of the computation load that can be saved. Thus, with larger storage load $r$, the saving of computation load to achieve $L^*(r)$ and $D^*(r)$ is larger. It will be clear later that the saving on the computation load is due to the fact that, under the assumption that each not computes all IVs it can computes, some of the IVs computed are not used in neither generating the signal, nor in the decoding process. 

The projections of the Pareto-optimal surface $\mathcal{O}$ into the uplink and downlink SCC space  correspond to the surfaces
\begin{IEEEeqnarray}{c}
\mathcal{O}^{\rm u}\triangleq\{P_{i-1}^{\rm u}P_i^{\rm u}P_K^{\rm u}: i\in[2:K-1]\}\notag
\end{IEEEeqnarray}
and
\begin{IEEEeqnarray}{c}
\mathcal{O}^{\rm d}\triangleq\{P_{i-1}^{\rm d}P_i^{\rm d}P_K^{\rm d}: i\in[2:K-1]\}, \notag
\end{IEEEeqnarray}
respectively. Observe that, for a given feasible $(r,c)$ pair, the optimal  upload is strictly larger than the optimal download.  We will see that this is achieved by performing some simple chain coding at the AP to combine the signals from different nodes. Interestingly, both the upload and download can be simultaneously achieved for a fixed $(r,c)$ pair.

%

\begin{remark}[Relation to Results in {\cite{Yan2022tradeoff}}]\label{remark2} One can observe that, the surfaces composing $L^*(r,c)$ and $\mathcal{O}^{\rm u}$ concide with the optimal communication load and the Pareto-optimal SCC tradeoff surface in the setup where the nodes directly connect to each other through a shared link (c.f. \cite[Fig. 2]{Yan2022tradeoff}), respectively.  It was showed in \cite{Yan2022tradeoff}, by dropping the computations of the IVs that are not used in the CDC scheme \cite{Li2018Tradeoff} as in \cite{Fragouli}, the resultant coded computing scheme can achieve the corner points of the Pareto-optimal SCC tradeoff surface (which is same as $\mathcal{O}^{\rm u}$). In fact, in our proposed scheme, each node performs the same map procedures as in \cite{Fragouli}, but the signals are sent to the AP. The AP performs a simple chain coding on the received signals to further compress the length of the signals, which leads to a further decrease of the download compared to the upload. We will present the whole process in Section \ref{sec:achieve}.
\end{remark}


%
%

\section{Achievability}\label{sec:achieve}

Since the set $\mathcal{O}$ is exactly all the Pareto-optimal points of the set $\mathcal{R}$ (Appendix \ref{App:OR}), we only need to prove the achievability of the hypersurface $\mathcal{O}$.
We will derive a coded computing scheme that achieves the SCC quadruple $P_i$. 
Moreover, for any fixed $\theta_1,\theta_2,\theta_3\in[0,1]$ such that $\theta_1+\theta_2+\theta_3=1$, divide the $N$ files into three groups of sizes\footnote{This requires that $\theta_1,\theta_2,\theta_3$ have to be rational. If any one is irrational, one can replace it by a rational number arbitrarily close to it. } $\theta_1N$, $\theta_2N$ and $\theta_3N$. By applying the scheme achieving the points $P_{i-1}$, $P_i$ and $P_K$ on the three groups of files, the resultant scheme achieves the point $P=\theta_1P_{i-1}+\theta_2P_i+\theta P_K$.  
Thus, we only need to prove the achievability of $P_i$, $i\in[K]$.

\subsection{Coded Distributed Computing for Star Network}
We now describe the scheme achieving $P_i$ for a fixed $i\in[K]$.  

Define
\begin{IEEEeqnarray}{c}
\mathbf{\Omega}_i\triangleq\left\{\mathcal{T}\subseteq [K]:|\mathcal{T}|=i\right\},\quad\forall\, i\in[K]. \notag
\end{IEEEeqnarray}

For $i=K$, $P_{K}=(K,1,0,0)$ is trivial, since each node can simply store all the files and computes their IVs as well as their reduce functions locally, with no communication loads. 

Consider a fixed $i\in[K-1]$,  the $N$ files are partitioned into
${K\choose i}$ batches, each containing
\begin{IEEEeqnarray}{c}
\eta_i=\frac{N}{{K\choose i}}\label{eqn:eta}
\end{IEEEeqnarray}
files. Each batch is then associated with a subset $\mathcal{T}$ of $[K]$ of cardinality $i$, i.e., an element in $\mathbf{\Omega}_i$. Let $\mathcal{W}_{\mathcal{T}}$ denote the batch of the $\eta_i$ files associated with set $\mathcal{T}$. Then,
  \begin{IEEEeqnarray}{c}
\mathcal{W}=\{w_1,\ldots,w_N\}=\bigcup_{\mathcal{T}\in\mathbf{\Omega}_i}\mathcal{W}_{\mathcal{T}}.\notag
\end{IEEEeqnarray}
Further let $\mathcal{U}_{\mathcal{T},k}$ be the set of IVs for output function $\phi_k$ that can be computed from the files in $\mathcal{W}_{\mathcal{T}}$:
\begin{IEEEeqnarray}{c}
\mathcal{U}_{\mathcal{T},k}\triangleq\{v_{k,n}:w_n\in\mathcal{W}_{\mathcal{T}}\}.\notag
\end{IEEEeqnarray}

We now describe the map, shuffle, and reduce procedures.

\begin{enumerate}
  \item \emph{Map Phase:} Each node $k$ stores
\begin{IEEEeqnarray}{c}
\mathcal{M}_k=\bigcup_{\mathcal{T}\in\mathbf{\Omega}_i:k\in\mathcal{T} }\mathcal{W}_{\mathcal{T}}, \notag
\end{IEEEeqnarray}
and computes the IVs
\begin{IEEEeqnarray}{c}
\mathcal{C}_k=\mathcal{C}_k^1\cup\mathcal{C}_k^2,\label{eqn:Ck}
\end{IEEEeqnarray}
where
\begin{subequations}\label{eqn:Ck12}
\begin{IEEEeqnarray}{rCl}
\mathcal{C}_k^1&=&\bigcup_{\mathcal{T}\in\mathbf{\Omega}_i:  k\in\mathcal{T}} \mathcal{U}_{\mathcal{T},k},\label{eqn:Ck1}\\
\mathcal{C}_k^2&=&\bigcup_{\mathcal{T}\in\mathbf{\Omega}_i:  k\in\mathcal{T}}\bigcup_{q\in \mathcal{K}\backslash \mathcal{T}} \mathcal{U}_{\mathcal{T},q}.\label{eqn:Ck2}
\end{IEEEeqnarray}
\end{subequations}
\quad In other words, for each batch $\mathcal{T}$, each node $k$ computes all
the IVs for its own function $k$, and all the IVs for the function $q$ if
node $q$ does not have the batch $\mathcal{T}$.

  \item \emph{Shuffle Phase:} For each element
    $\mathcal{T}\in\mathbf{\Omega}_i$ and each index $j\in \mathcal{K}
    \backslash \mathcal{T}$, we partition
 the set $\mathcal{U}_{\mathcal{T},j}$ into $i$ smaller subsets
 \begin{IEEEeqnarray}{c}
\mathcal{U}_{\mathcal{T},j}= \left\{ \mathcal{U}_{\mathcal{T},j}^k\colon \; k \in \mathcal{T}\right\}\label{eqn:Upartition}
 \end{IEEEeqnarray} of equal size.

\quad In the \emph{upload sub-phase},  for each  $\mathcal{S}\in\mathbf{\Omega}_{i+1}$ and $k\in \mathcal{S}$, by \eqref{eqn:Ck2},  node $k$ can compute
the signal
\begin{IEEEeqnarray}{c}
X_{ \mathcal{S}}^{k}\triangleq\bigoplus_{l\in \mathcal{S}\backslash\{k\}}\mathcal{U}_{\mathcal{S}\backslash\{l\},l}^k\notag
\end{IEEEeqnarray}
from the IVs calculated during the map phase.  Node $k$ thus sends  the multicast signal
 \begin{IEEEeqnarray}{rCl}
 	X_k = \left\{ X_{ \mathcal{S}}^{k} \colon \; \mathcal{S} \in \mathbf{\Omega}_{i+1} \; \textnormal{such that } k\in  \mathcal{S} \right\}\notag
 \end{IEEEeqnarray}
to the AP $R$. Thus, the AP $R$ receives the signals $X_1,\ldots,X_K$.

\quad In the \emph{download sub-phase}, for each $\mathcal{S}=\{k_1,\ldots,k_{i+1}\}\in\mathbf{\Omega}_{i+1}$, the AP $R$ creates a signal, 
\begin{IEEEeqnarray}{rCl}
X_{\mathcal{S}}&\triangleq &(X_{\mathcal{S}}^{k_1}\oplus X_{\mathcal{S}}^{k_2}, X_{\mathcal{S}}^{k_2}\oplus X_{\mathcal{S}}^{k_3},\ldots,X_{\mathcal{S}}^{k_i}\oplus X_{\mathcal{S}}^{k_{i+1}}). \IEEEeqnarraynumspace\label{eqn:XS}
\end{IEEEeqnarray}

\quad Then the AP broadcast the signal
\begin{IEEEeqnarray}{rCl}
X\triangleq\{X_{\mathcal{S}}:\mathcal{S}\in\mathbf{\Omega}_{i+1}\}.\label{eq:relayX}
\end{IEEEeqnarray}
  \item \emph{Reduce Phase:} Notice that $\mathcal{C}_k^{2}$ only
    contains the IVs $v_{q,n}$ where $q \neq k$. Thus, by
    \eqref{eqn:Ck} and \eqref{eqn:Ck1},  during the shuffle phase
    each node $k$ needs to learn  all the IVs in
\begin{IEEEeqnarray}{c}
	\bigcup_{\substack{\mathcal{T}\in \mathbf{\Omega}_i \colon  k\notin \mathcal{T}} }  \mathcal{U}_{\mathcal{T},k}.\notag
\end{IEEEeqnarray}

\quad Fix an arbitrary  $\mathcal{T}\in \mathbf{\Omega}_{i}$ such that $k \notin \mathcal{T}$. From the received multicast message $X_{\mathcal{T}\cup\{k\}}$, since the signal $X_{\mathcal{T}\cup \{k\}}^k$ is generated by node $k$,  by \eqref{eqn:XS}, node $k$ can decode $X_{\mathcal{T}\cup \{k\}}^j$ for all $j\in\mathcal{T}$, where the signal
\begin{IEEEeqnarray}{c}
X_{\mathcal{T}\cup \{k\}}^j=\bigoplus_{l\in\mathcal{T}\cup \{k\}\backslash\{j\}}\mathcal{U}_{\mathcal{T}\cup \{k\}\backslash\{l\},l}^j\notag
\end{IEEEeqnarray}
is sent by node $j$ during the shuffle phase. 
For any fixed $j\in\mathcal{T}$, node $k$ can recover the missing IV   $\mathcal{U}_{\mathcal{T},k}^j$ through a simple XOR operation:
\begin{IEEEeqnarray}{c}
\mathcal{U}_{\mathcal{T},k}^j=X_{\mathcal{T}\cup \{k\}}^j\oplus \bigoplus_{l\in\mathcal{T}\backslash\{j\}}\mathcal{U}_{\mathcal{T}\cup \{k\}\backslash\{l\},l}^j,\label{eqn:decodeU}
\end{IEEEeqnarray}
where $\mathcal{U}_{\mathcal{T}\cup \{k\}\backslash\{l\},l}^j$ is calculated at node $k$ by \eqref{eqn:Ck2} and \eqref{eqn:Upartition} for all $l\in\mathcal{T}\backslash\{j\}$.  Moreover, node $k$ can decode $\mathcal{U}_{\mathcal{T},k}$ from
\begin{IEEEeqnarray}{c}
\left\{X_{\mathcal{T}\cup \{k\}}^{j}:j\in\mathcal{T}\right\}.\notag
\end{IEEEeqnarray}
by \eqref{eqn:Upartition} and \eqref{eqn:decodeU}. 
 After collecting all the missing IVs, node $k$ can proceed to compute the reduce function \eqref{eqn:reducefunction}.
\end{enumerate}
\begin{remark} [Comparison with \cite{Li2017Framework}] Compared to the coded computing scheme in \cite{Li2017Framework}, two differences of the above scheme are:
\begin{enumerate}
\item In the map phase, each node only needs to compute the IVs described in \eqref{eqn:Ck} and \eqref{eqn:Ck12}, because only those IVs are useful for creating or decoding the coded signals,  while in \cite{Li2017Framework}, all the IVs pertaining to the files in $\mathcal{M}_k$ are computed, i.e., node $k$ computes 
\begin{IEEEeqnarray}{c}
\widetilde{\mathcal{C}}\triangleq \bigcup_{\mathcal{T}\in\mathbf{\Omega}_i, k\in\mathcal{T}}\bigcup_{q\in[K]}\mathcal{U}_{\mathcal{T},q}.
\end{IEEEeqnarray}
This scheme in fact achieves the point $Q_i$, which is inferior to $P_i$ for $i>0$ in computation load. The idea of removing the redundancy has been proposed in the setup where the nodes connects to each other directly through a bus link
 by  Ezzeldin \cite{Fragouli} and Yan \emph{et al} \cite{Yan2022tradeoff}.  
\item In \eqref{eqn:XS}, for any node set $\mathcal{S}$ of size $i+1$, we used a simple chain coding on the signals  to form $i$ signals, while in \cite{Li2017Framework}, it uses  random coding on the signals $\{X_{\mathcal{S}}^{k}:k\in\mathcal{S}\}$ to form $i$ coded signals. The advantage of chain coding in \eqref{eqn:XS} is obvious: 
\begin{enumerate}
\item It has smaller encoding and decoding complexities;
\item It can be operated on the binary field $\mathbb{F}_2$;
\item The order of nodes in the chain can be arbitrary.  It makes sense in some scenarios: the signals $\{X_{\mathcal{S}}^{k}:k\in\mathcal{S}\}$ may arrive at different time points.  Consider the case that the signals arrive in the ordder $X_{\mathcal{S}}^{k_1},X_{\mathcal{S}}^{k_2},\ldots,X_{\mathcal{S}}^{k_{i+1}}$, to perform the encoding \eqref{eqn:XS}, at any time the AP only needs to keep one signal in its buffer, because each coordinate in \eqref{eqn:XS} only depends on two consecutive signals. While with random linear coding, the AP typically have to wait for all signals $\{X_{\mathcal{S}}^{k}:k\in\mathcal{S}\}$. Thus, the chain coding can reduce the buffer size at the AP and the node to node delay.
\end{enumerate}
\end{enumerate}
\end{remark}
\begin{remark}[PDA framework]  In \cite{Yan2022tradeoff,YanPDAITW2018,KumarPDA:ISIT}, a coded computing scheme was derived based on \emph{placement delivery array (PDA)}, which was proposed in \cite{Yan2017PDA} to explore coded caching schemes with uncoded placement \cite{Maddah2014Fundamental}. In particular, it turns out that the Maddah-Ali and Niesen's coded caching scheme corresponds to a special structure of PDA (referred to as MAN-PDA).  It was showed in  \cite{Yan2022tradeoff} that, with any given PDA belonging to a special class (defined as \emph{PDA for distributed computing (Comp-PDA)}),   one can always obtain a coded computing scheme. The class of PDAs achieving the Pareto-optimal tradeoff surface was characterized in \cite{Yan2022tradeoff}. The advantage of establishing the PDA framework is, various known PDA structure, e.g., the constructions in \cite{Yan2017PDA, Shangguan2018Hyper,Yan2018bipartite} can be directly utilized to obtain coded computing schemes with low \emph{file complexity}\footnote{The file complexity of a coded computing scheme is defined as the smallest number of files required to implement the scheme, e.g., the file complexity of the proposed scheme achiving $P_i$ is ${K\choose i}$.}. In our setup, similar connections between coded computing schemes and Comp-PDA can be  established, by following the same steps as in \cite{Yan2022tradeoff} for upload singals, and 
incoporating the chain coding \eqref{eqn:XS} on all multicast signals from the Comp-PDA for the downlink signals. For example, the scheme described in Fig. \ref{fig:example} can be derived from the PDA
\begin{IEEEeqnarray}{c}
\left[\begin{array}{ccc}*&1&*\\
1&*&*\\
*&*&1
\end{array}\right],\notag
\end{IEEEeqnarray}
for details of forming the upload signals in Fig. \ref{fig:a}, one can refer to \cite[Example 4]{Yan2022tradeoff}.
\end{remark}

\subsection{Performance Analysis} 
We analyze the performance of  the scheme.
\begin{enumerate}
  \item \emph{Storage Load:} The number of batches in $\mathcal{M}_k$ is
$
{K-1\choose i-1}
$,
each consisting of  $\eta_i $ files. Thus, the  storage load is
\begin{IEEEeqnarray}{rCl}
r=\frac{1}{N}\cdot K\cdot{K-1\choose i-1}\cdot\eta_i=i.\label{eqn:storage}
\end{IEEEeqnarray}

  \item\emph{Computation Load:} 
  Since $\mathcal{C}_k^1\cap \mathcal{C}_k^2=\emptyset$, we have
  $|\mathcal{C}_k|=|\mathcal{C}_{k}^1|+|\mathcal{C}_k^2|$.
From \eqref{eqn:eta}, \eqref{eqn:Ck1}, and \eqref{eqn:Ck2}, we have
\begin{IEEEeqnarray}{rCl}
|\mathcal{C}_{k}^1|&=&{K-1\choose i-1} \cdot\eta_i=\frac{iN}{K},\notag\\
|\mathcal{C}_{k}^2|&=&{K-1\choose i-1}\cdot(K-i)\cdot\eta_i\notag\\
&=&\left(1-\frac{i}{K}\right)\cdot i\cdot N.\notag
\end{IEEEeqnarray}
Thus,  the computation load is
\begin{IEEEeqnarray}{c}
c=\frac{\sum_{k=1}^K|\mathcal{C}_k|}{NK}
=i\left(1-\frac{i-1}{K}\right).\label{eqn:computation}
\end{IEEEeqnarray}
  \item \emph{Communication Load:} The number of signals that each node $k$ transmits is
${K-1\choose i}$,
each of size $\frac{\eta_i\cdot V}{i}$ bits. Thus, the length of the signal $X_k$ is $l_k={K-1\choose i}\frac{\eta_i\cdot V}{i}$ bits. Therefore,
the \emph{upload} is
\begin{IEEEeqnarray}{c}
L=\frac{\sum_{k=1}^Kl_k}{NKV}
=\frac{1}{i}\cdot\left(1-\frac{i}{K}\right).\label{eqn:communication}
\end{IEEEeqnarray}
By \eqref{eqn:XS} and \eqref{eq:relayX}, the AP $R$ transmits ${K\choose i+1}\cdot i$ signals, each of size $\frac{\eta_i\cdot V}{i}$ bits, thus the \emph{download} is
\begin{IEEEeqnarray}{rCl}
D&=&\frac{1}{NKV}\cdot{K\choose i+1}\cdot i\cdot \frac{\eta_i\cdot V}{i}=\frac{1}{i+1}\left(1-\frac{i}{K}\right). \label{eqn:downlink:communication}
\end{IEEEeqnarray}
\end{enumerate}

From \eqref{eqn:storage}, \eqref{eqn:computation}, 
\eqref{eqn:communication} and \eqref{eqn:downlink:communication}, we show the achievability of the SCC quadruple $P_i$.




\section{Converse}\label{sec:converse}
We need to prove that for any achievable $(r,c,L,D)$ satisfying \eqref{eq:interesting}, 
\begin{subequations}\label{lowerbound}
\begin{IEEEeqnarray}{rCl}
L&\geq & L^*(r,c),\label{lowerbound:a}\\
 D&\geq & D^*(r,c). \label{lowerbound:b}
\end{IEEEeqnarray}
\end{subequations}
Consider a coded distributed computing scheme achieving $(r,c,L,D)$,  with file allocations $\mathcal{M}_{[K]}$, IV allocations $\mathcal{C}_{[K]}$, uplink signals $X_{[K]}$ and downlink signal $X$. By the decoding condition \eqref{eqn:adjust1b}, 
\begin{IEEEeqnarray}{c}
H(\mathcal{V}_k|X,\mathcal{C}_k)=0,\quad\forall\,k\in[K]. \notag
\end{IEEEeqnarray}
Thus for any $k\in[K]$,
\begin{IEEEeqnarray}{rCl}
&&H(\mathcal{V}_k|X_1,\ldots,X_K,\mathcal{C}_k)\notag\\
&\overset{(a)}{=}&H(\mathcal{V}_k|X_1,\ldots,X_K,X,\mathcal{C}_k)\notag\\
&\leq& H(\mathcal{V}_k|X,\mathcal{C}_k)\notag\\
&=&0,\notag
\end{IEEEeqnarray}
where $(a)$ follows since the downlink signal $X$ is determined by the uplink signals $X_{[K]}$ by \eqref{eqn:X}. 

That is, with the signals $X_1,\ldots, X_K$ and the locally computed IVs $\mathcal{C}_k$, node $k$ can decode all the IVs it needs. As a result, the file allocations $\mathcal{M}_{[K]}$, IV allocations $\mathcal{C}_{[K]}$ and the uplink singals $X_{[K]}$ consisitute an valid scheme for the distributed computing system where the nodes are connected through a bus shared link directly, as investigated in   \cite{Yan2022tradeoff}. Therefore, by the results in \cite[Theorem 2]{Yan2022tradeoff}, we have proved \eqref{lowerbound:a}. 

We proceed to prove the \eqref{lowerbound:b}. 
For any $k\in[K]$ and nonempty $\mathcal{S}\subseteq[K]\backslash\{k\}$, define
\begin{IEEEeqnarray}{rCl}
\mathcal{B}_{k,\mathcal{S}}&\triangleq&\{v_{k,n}: v_{k,n}~\mbox{is exclusively computed} \mbox{~by the nodes in}~\mathcal{S}\},\notag\\
\widetilde{\mathcal{B}}_k&\triangleq&\{v_{k,n}: v_{k,n}~\mbox{is the computed by node}~k\}.\notag
\end{IEEEeqnarray}
Let $b_{k,\mathcal{S}}$ be the cardinality of the set $\mathcal{B}_{k,\mathcal{S}}$ and $\tilde{b}_k$ be the cardinality of $\widetilde{B}_k$.  Obviously, the subsets $\{\mathcal{B}_{k,\mathcal{S}}:\mathcal{S}\subseteq[K]\backslash\{k\},\mathcal{S}\neq \emptyset\}$ and $\widetilde{\mathcal{B}}_k$ form a partition of the IVs $\mathcal{V}_k$, thus
\begin{IEEEeqnarray}{c}
\tilde{b}_k+\sum_{\mathcal{S}\subseteq[K],\mathcal{S}\neq\emptyset} b_{k,\mathcal{S}}=N.\notag
\end{IEEEeqnarray}
For each $j\in[K-1]$, the set of IVs not computed locally but exclusively computed by $j$ other nodes are
\begin{IEEEeqnarray}{c}
\mathcal{B}_j=\bigcup_{k\in[K]}\bigcup_{\mathcal{S}\subseteq[K]\backslash\{k\}, |\mathcal{S}|=j} \mathcal{B}_{k,\mathcal{S}}. \notag
\end{IEEEeqnarray}
Then the cardinality of set $\mathcal{B}_j$ is given by 
\begin{IEEEeqnarray}{c}
b_j\triangleq \sum_{k\in[K]}\sum_{\mathcal{S}\subseteq[K]\backslash\{k\},|\mathcal{S}|=j}b_{k,\mathcal{S}},~~\forall j\in[K-1].\IEEEeqnarraynumspace\label{eqn:bjbkS}
\end{IEEEeqnarray}

To prove the lower bound in \eqref{lowerbound:b}, we need the following two lemmas. 
\begin{lemma}\label{lemma1} The entropy of the download signal $X$ satisfy
\begin{IEEEeqnarray}{c}
H(X)\geq V\sum_{j=1}^{K-1}\frac{b_j}{j+1}.\notag
\end{IEEEeqnarray}
\end{lemma}
\begin{IEEEproof} 
Assume that the AP holds all IVs $\mathcal{V}_k$, then the access  point can create the signal $X$. Consider the data exchange problem\footnote{Data exchange problem was defined in \cite{Krishnan2020}, where each of the nodes holds a subset of the information bits, and request another subset of information bits. } formed by the AP and the $K$ nodes, where only the AP sends the signal $X$ to all the $K$ nodes. Notice that, in this system, each bits in $\mathcal{B}_{k,\mathcal{S}}$ is cached at the AP and the nodes in $\mathcal{S}$, but only demanded by node $k$. Thus, by the lower bound in \cite[Theorem 1]{Krishnan2020}, 
\begin{IEEEeqnarray}{rCl}
H(X)
&\geq&V\sum_{k\in[K]}\sum_{\mathcal{S}\subseteq[K]\backslash\{k\}}\frac{1}{(|\mathcal{S}|+1)+1-1}\cdot b_{k,\mathcal{S}}\notag\\
&=&V\sum_{k=1}^K\sum_{j=1}^{K-1}\sum_{\mathcal{S}\subseteq[K]\backslash\{k\},|\mathcal{S}|=j}\frac{1}{j+1}\cdot b_{k,\mathcal{S}}\notag\\
&\overset{(a)}{=}&V\sum_{j=1}^{K-1}\frac{1}{j+1}\sum_{k=1}^K\sum_{\mathcal{S}\subseteq[K]\backslash\{k\},|\mathcal{S}|=j} b_{k,\mathcal{S}}\notag\\
&=&V\sum_{j=1}^{K-1}\frac{b_j}{j+1},\notag
\end{IEEEeqnarray} 
where in $(a)$, we utilized \eqref{eqn:bjbkS}. 
\end{IEEEproof}

The following lemma was proved in \cite[Lemma 2]{Yan2022tradeoff}.
\begin{lemma}\label{lemma2}The parameters $b_1,\ldots,b_{K-1}$ defined in \eqref{eqn:bjbkS} satisfy
\begin{IEEEeqnarray}{rCl}
\sum_{j=1}^{K-1}b_j\geq N(K-r),\notag\\
\sum_{j=1}^{K-1}(j-1)b_j\leq (c-1)NK. \notag
\end{IEEEeqnarray}
\end{lemma}

For a fixed $r\in[1:K]$, and each $i\in[K]$, define
\begin{IEEEeqnarray}{c}
c_i\triangleq1+\Big(1-\frac{r}{K}\Big)(i-1).\label{eqn:c_i}
\end{IEEEeqnarray}
Let $\lambda_i,\mu_i\in\mathbb{R}^+$ such that
\begin{subequations}
\begin{IEEEeqnarray}{rCl}
\lambda_ix+\mu_i|_{x=c_{i-1}}&=&\frac{1}{c_{i-1}+1-2r/K}\cdot\Big(1-\frac{r}{K}\Big)^2\notag\\
&=&\frac{1}{i}\Big(1-\frac{r}{K}\Big),\label{lab:a}\\
\lambda_ix+\mu_i|_{x=c_i}&=&\frac{1}{c_i+1-2r/K}\cdot\Big(1-\frac{r}{K}\Big)^2\notag\\
&=&\frac{1}{i+1}\Big(1-\frac{r}{K}\Big).\label{lab:b}
\end{IEEEeqnarray} 
\end{subequations}
From \eqref{lab:a} and \eqref{lab:b}, the following relationships hold:
\begin{subequations}\label{eqn:signs}
\begin{IEEEeqnarray}{rCl}
\lambda_i&=&-\frac{1}{i(i+1)}<0,\label{eqn:lab:i}\\
\mu_i&=&\frac{2i-1}{i(i+1)}\Big(1-\frac{r}{K}\Big)+\frac{1}{i(i+1)}>0,\IEEEeqnarraynumspace\label{eqn:mu:i}\\
\lambda_i+\mu_i&=&\frac{2i-1}{i(i+1)}\Big(1-\frac{r}{K}\Big)>0.\label{eqn:lab:plus:mu}
\end{IEEEeqnarray}
\end{subequations}
Moreover, by its convexity over $x\in[1,\infty)$, the function 
\begin{IEEEeqnarray}{c}
\frac{1}{x+1-2r/K}\Big(1-\frac{r}{K}\Big)^2-(\lambda_ix+\mu_i)\notag
\end{IEEEeqnarray}
must be nonnegative outside the interval formed by the two zero points, i.e., 
\begin{IEEEeqnarray}{c}
\frac{1}{x+1-2r/K}\Big(1-\frac{r}{K}\Big)^2\geq\lambda_ix+\mu_i,\quad \forall~x\in[1,c_{i-1}]\cup[c_i,\infty).\notag
\end{IEEEeqnarray}
Therefore, 
\begin{IEEEeqnarray}{c}
\frac{1}{c_j+1-2r/K}\Big(1-\frac{r}{K}\Big)^2\geq \lambda_ic_j+\mu_i,\quad\forall~j\in[K-1].\label{eqn:inqcj}
\end{IEEEeqnarray}
Now, we are ready to derive the lower bound for the download $D$:
\begin{IEEEeqnarray}{rCl}
D&\geq& \frac{H(X)}{NKV}\notag\\
&\overset{(a)}{\geq}&\sum_{j=1}^{K-1}\frac{b_j}{NK}\cdot\frac{1}{j+1}\notag\\
&\overset{(b)}{=}&\frac{1}{N(K-r)}\sum_{j=1}^{K-1}b_j\cdot\frac{1}{c_j+1-2r/K}\Big(1-\frac{r}{K}\Big)^2\notag\\
&\overset{(c)}{\geq}&\frac{1}{N(K-r)}\sum_{j=1}^{K-1}b_j(\lambda_ic_j+\mu_i)\notag\\
&\overset{(d)}{=}&\frac{1}{N(K-r)}\sum_{j=1}^{K-1}b_j\Big(\lambda_i\Big(1+\Big(1-\frac{r}{K}\Big)(j-1)\Big)+\mu_i\Big)\notag\\
&=&\frac{\lambda_i}{NK}\cdot\sum_{j=1}^{K-1}(j-1)b_j+\frac{\lambda_i+\mu_i}{N(K-r)}\cdot\sum_{j=1}^{K-1}b_j\notag\\
&\overset{(e)}{\geq}&\frac{\lambda_i}{NK}\cdot(c-1)NK+\frac{\lambda_i+\mu_i}{N(K-r)}\cdot N(K-r)\notag\\
&=&\lambda_ic+\mu_i\notag\\
&=&-\frac{2i-1}{Ki(i+1)}r-\frac{1}{i(i+1)}c+\frac{2}{i+1}.\label{eq:tri} 
\end{IEEEeqnarray}
where $(a)$ follows from Lemma \ref{lemma1}; $(b)$ and $(d)$ follow from the definition of $c_i$ in \eqref{eqn:c_i}; $(c)$ follows from \eqref{eqn:inqcj}; and $(e)$ follows from Lemma \ref{lemma2} and the signs of $\lambda_i$ and $\lambda_i+\mu_i$ in \eqref{eqn:signs}.

Notice that the three points $P_{i-1}^{\rm d}, P_{i}^{\rm d}$ and $P_K^{\rm d}$ defined in \eqref{def:Pd} satisfy \eqref{eq:tri} with equality. Thus, the inequalities above indicate that all the feasible points $(r,c,L,D)$ must satisfy that the projection into the download SCC space $(r,c,D)$ must lie above the plane containing $\triangle P_{i-1}^{\rm d}P_i^{\rm d}P_K^{\rm d}$. 

Furthmore, $D$ should be lower bounded by the optimal download even if  each node computes all the IVs that can be computed locally from their stored file, i.e., a similar setup as in \cite{Li2017Framework}. The converse in \cite{Li2017Framework} indicates that $L$ is lower bounded as follows in the $r$-$D$ plane\footnote{Although the setup in \cite{Li2017Framework} assumes a fixed storage capacity at each node, the proof the  following inequality do not rely on this assumption.}:
\begin{IEEEeqnarray}{c}
D\geq {\rm{Conv}}\left(\frac{1}{r}\Big(1-\frac{r}{K}\Big)\right),~~r\in\{1,2,\ldots,K\}.\IEEEeqnarraynumspace \label{eq:ConvD}
\end{IEEEeqnarray}
Finally, by the lower bounds in \eqref{eq:tri} for $i=2,3,\ldots,K-1$ and \eqref{eq:ConvD}, $D$ is lower bounded by $D^*(r,c)$, i.e., the lower bound \eqref{lowerbound:b} is proved. 

\section{Conclusion}\label{sec:conclusion}
In this paper, the Pareto-optimal storage-computation-upload-download tradeoff surface is characterized for the MapReduce distributed computing system, where the nodes have to exchage intermediate values through an access  point that can broadcast signals to all nodes. It turns out that, for a given storage-computation pair $(r,c)$, the optimal upload and download can be simultaneously achieved.  Information-theoretical bounds matching the achievable communication load are provided for both uplink and downlink.

%

\appendices
\section{The Relation of Hypersurface $\mathcal{O}$ and Region $\mathcal{R}$}\label{App:OR}
\begin{figure}[t]
 	\centering
 	\includegraphics[width=0.5\textwidth]{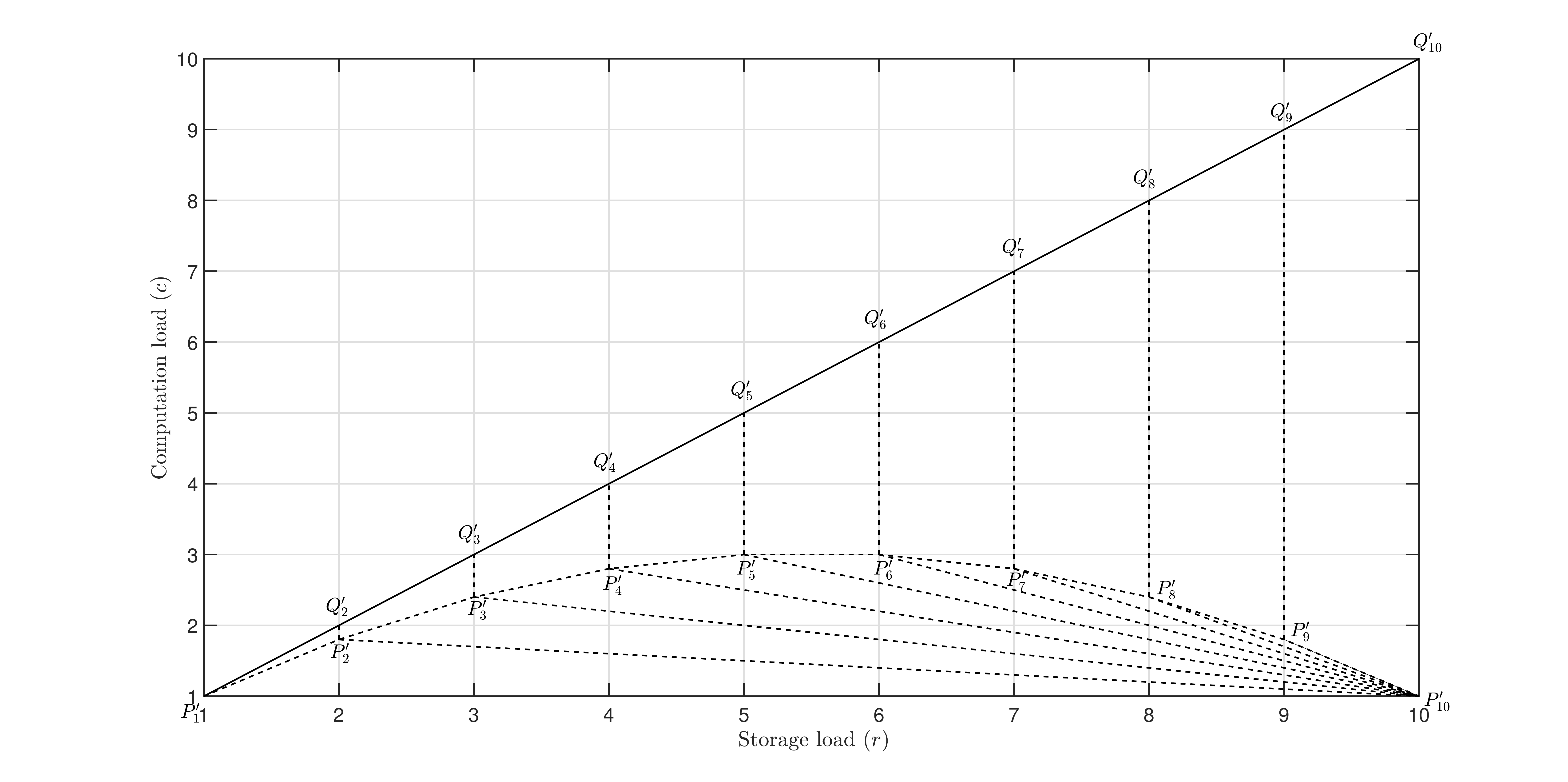}
 	\caption{The projections of $P_i$ and $Q_i$ $(i\in[K])$ to the storage-computation subspace ($r$-$c$ plane).
 		 }\label{fig:rcplane}
 \end{figure}

%
We now prove that $\mathcal{O}$ is the Pareto-optimal surface of the region $\mathcal{R}$.  Obviously, all Pareto-optimal  points must lie on the surface
\begin{IEEEeqnarray}{c}
\mathcal{F}=\{(r,c,L^*(r,c),D^*(r,c)):1\leq c\leq r\leq K\}. \notag
\end{IEEEeqnarray}
Let the projections of points $P_i$ and $Q_i$ into the $r$-$c$ plane be $P_i'$ and $Q_i'$ ($i\in[K]$), respectively\footnote{Notice that the projections of $P_i^{\rm u},Q_i^{\rm u}$ and $P_i^{\rm d}, Q_i^{\rm d}$ into the $r$-$c$ plane are the same as the ones  of the points $P_i$ and $Q_i$. As a result, the projections of $\triangle P_1^{\rm u}P_2^{\rm u}Q_2^{\rm u}$/$\triangle P_1^{\rm d}P_2^{\rm d}Q_2^{\rm d}$, $\triangle P_{i-1}^{\rm u}P_{i}^{\rm u}P_K^{\rm u}$/$\triangle P_{i-1}^{\rm u}P_{i}^{\rm u}P_K^{\rm u}$,  and $\boxminus P_i^{\rm u}Q_i^{\rm u}Q_{i+1}^{\rm u}P_{i+1}^{\rm u}$/$\boxminus P_i^{\rm d}Q_i^{\rm d}Q_{i+1}^{\rm d}P_{i+1}^{\rm d}$ into the $r$-$c$ plane are $\triangle P_1'P_2'Q_2'$, $\triangle P_{i-1}'P_i'P_K'$  and $\boxminus P_i'Q_i'Q_{i+1}'P_{i+1}'$, respectively.}, i.e., 
\begin{IEEEeqnarray}{c}
P_i'=\Big(i,i\Big(1-\frac{i-1}{K}\Big)\Big),\quad Q_i'=(i,i).\notag
\end{IEEEeqnarray}
Let the projection of the surface $\mathcal{F}$ to the $r$-$c$ plane be 
\begin{IEEEeqnarray}{c}
\mathcal{F}'\triangleq\{(r,c):1\leq c\leq r\leq K\}=\triangle P_1'P_K'Q_K'.\notag
\end{IEEEeqnarray}
 Notice that, here the ``projection" map is one-to-one.  Moreover, $\mathcal{F}'$ can be decomposed into (see Fig. \ref{fig:rcplane}) 
\begin{IEEEeqnarray}{rCl}
\mathcal{F}'&=&\triangle P_1'P_2'Q_2'\cup \mathop\cup_{i=2}^{K-1} \triangle P_{i-1}'P_i'P_K'\cup\mathop\cup_{i=2}^{K-1}\boxminus P_i'Q_i'Q_{i+1}'P_{i+1}' .\notag
\end{IEEEeqnarray}
Since the triangle $\triangle P_1^{\rm u}P_2^{\rm u}Q_2^{\rm u}$ and the trapezoids $\boxminus P_i^{\rm u}Q_i^{\rm u}Q_{i+1}^{\rm u}P_{i+1}^{\rm u}$ in the uplink SCC space ($i\in[2:K-1]$) are parallel to $c$-axis, and so as the triangle $\triangle P_1^{\rm d}P_2^{\rm d}Q_2^{\rm d}$ and the trapezoids $\boxminus P_i^{\rm d}Q_i^{\rm d}Q_{i+1}^{\rm d}P_{i+1}^{\rm d}$ in the downlink SCC space, all the points $(r,c,L^*(r,c),D^*(r,c))\in\mathcal{F}$ such that 
\begin{IEEEeqnarray}{rCl}
&(r,c)\in&\triangle P_1'P_2'Q_2' \cup\mathop\cup_{i=2}^{K-1}\boxminus P_i'Q_i'Q_{i+1}'P_{i+1}'\backslash\mathop\cup_{i=2}^{K-1}\triangle P_{i-1}'P_iP_K'\IEEEeqnarraynumspace\label{eq:parallel}
\end{IEEEeqnarray}
cannot be Pareto-optimal. In the following, we prove that, all the points $(r,c,L^*(r,c),D^*(r,c))$  such that 
\begin{IEEEeqnarray}{c}
(r,c)\in\mathop\cup_{i=2}^{K-1}\triangle P_{i-1}'P_i'P_K'\label{eq:pro:optimal}
\end{IEEEeqnarray}
are Pareto-optimal.

Now  fix a quadruple $(r_1,c_1,L^*(r_1,c_1),D^*(r_1,c_1))\in\mathcal{F}$ that satisfies \eqref{eq:pro:optimal}. We show that it is Pareto-optimal. To this end, consider any other triple $(r_2,c_2,L_2,D_2)\in\mathcal{R}$ that satisfies
\begin{subequations}\label{eq:rcLcomp}
\begin{IEEEeqnarray}{rClrCl}
r_2&\leq& r_1,&
c_2&\leq& c_1,\\
L_2&\leq& L^*(r_1,c_1),&\quad D_2&\leq& D^*(r_1,c_1).
\end{IEEEeqnarray}
\end{subequations}
We show by contradiction that all four inequalities must hold with equality. Notice that,  $(r_2,c_2)$  either satisfies   \eqref{eq:parallel} or \eqref{eq:pro:optimal}. 
\begin{enumerate}
  \item Assume that $(r_2,c_2)$  satisfies  \eqref{eq:pro:optimal}. If $r_2< r_1$ or $c_2<c_1$, then consider the uplink SCC subspace, one can verify that the points $P_{i-1}^{\rm u}$, $P_i^{\rm u}$ and $P_K^{\rm u}$ are on the surface
\begin{IEEEeqnarray}{c}
L=-\frac{1}{i(i-1)}c-\frac{2}{Ki}r+\frac{2i-1}{i(i-1)}.\label{eq:triangle}
\end{IEEEeqnarray}
Therefore, it must hold that 
    \begin{IEEEeqnarray}{c} \label{eq:Ldir}
L^*(r_2,c_2)> L^*(r_1,c_1),
  \end{IEEEeqnarray}
 because all the surfaces containing $\triangle P_{i-1}^{\rm u}P_i^{\rm u} P_K^{\rm u}$ ($i\in[2:K-1]$) have positive directional derIVtives along $(r_2-r_1,c_2-c_1)$  by \eqref{eq:triangle}. Since $(r_2,c_2,L_2,D_2)\in\mathcal{R}$, we have $L_2 \geq L^*(r_2,c_2)$ and thus by \eqref{eq:Ldir}, $L_2>L^*(r_1,c_1)$,
  which contradicts  \eqref{eq:rcLcomp}. 
Therefore, it must hold that $r_2=r_1$ and $c_2=c_1$. Then obviously, $L_2 \geq L^*(r_2,c_2)=L^*(r_1,c_1)$ and $D_2\geq D^*(r_2,c_2)=D^*(r_1,c_1)$, thus all equalities in \eqref{eq:rcLcomp} hold. 
  \item Assume now that $(r_2,c_2)$ satisfies \eqref{eq:parallel}. Then,  $(r_2,c_2)$ must lie on at least  one of the $K-1$ facets
   \begin{IEEEeqnarray}{c}
\triangle P_1'P_2'Q_2'~\mbox{or}~
 \boxminus P_i'Q_i'Q_{i+1}'P_{i+1}',~~ i\in[2:K-1], \notag
 \end{IEEEeqnarray}
and it must not lie on the line segments $P_{i-1}'P_i', i\in[2:K]$. As the facets $\triangle P_1^{\rm u}P_2^{\rm u}Q_2^{\rm u}$, $\boxminus P_i^{\rm u}Q_i^{\rm u}Q_{i+1}^{\rm u}P_{i+1}^{\rm u}$ ($i\in[2:K-1]$) in the uplink SCC subspace are all parellel to the $c$-axis, and so as the facets $\triangle P_1^{\rm d}P_2^{\rm d}Q_2^{\rm d}$, $\boxminus P_i^{\rm d}Q_i^{\rm d}Q_{i+1}^{\rm d}P_{i+1}^{\rm d}$ ($i\in[2:K-1]$) in the downlink SCC facets, there exists $c_3<c_2\leq c_1$ such that $(r_2,c_3)$ satisfies \eqref{eq:pro:optimal}, and 
\begin{IEEEeqnarray}{c}
L^*(r_2,c_3)=L^*(r_2,c_2), ~D^*(r_2,c_3)=D^*(r_2,c_2).\notag 
\end{IEEEeqnarray}
 Therefore,
      \begin{IEEEeqnarray}{rCl}
      L_2\geq  L^*(r_2,c_2)=L^*(r_2,c_3)\overset{(a)}{>}L^*(r_1,c_1),\IEEEeqnarraynumspace\label{eqn:largerthan}
      \end{IEEEeqnarray}
      where $(a)$ follows by proof step 1). 
      But \eqref{eqn:largerthan} contradicts with \eqref{eq:rcLcomp}.
\end{enumerate}

From the above analysis, we conclude that, the set of all Pareto-optimal points of $\mathcal{R}$ is exactly all the quadruples $(r,c,L^*(r,c),D^*(r,c))\in\mathcal{F}$ satisfying \eqref{eq:pro:optimal}. Notice that those points are exactly the surface $\mathcal{O}$ defined in \eqref{surfaceO}.

\ifCLASSOPTIONcaptionsoff
  \newpage
\fi

%

%
%
%





\end{document}